\documentclass[prb,twocolumn,showpacs,floatfix,amsfonts]{revtex4}
\usepackage{graphicx,graphics,color,epsfig}% Include figure files
\usepackage{bm}
\usepackage{amsmath}
\usepackage{amssymb}
\usepackage{epstopdf}
\begin{document}
\preprint{}
\title{Interplay between Superconductivity and Antiferromagnetism in a Multi-layered System}
\author{H. T. Quan and Jian-Xin Zhu }
\email{jxzhu@lanl.gov}
\homepage{http://theory.lanl.gov}
\affiliation{Theoretical Division, Los Alamos National Laboratory,
Los Alamos, New Mexico 87545, USA}

\begin{abstract}

Based on a microscopic model, we study the interplay between
superconductivity and antiferromagnetism in a multi-layered system,
where two superconductors are separated by an antiferromagnetic
region. Within a self-consistent mean-field theory, this system is
solved numerically. We find that the antiferromagnetism in the
middle layers profoundly affects the supercurrent flowing across the
junction, while the phase difference across  the junction influences
the development of antiferromagnetism in the middle layers. This
study may not only shed new light on % the mechanism for high-$T_{c}$ superconductors
material design and material engineering, but also bring important insights to building
Josephson-junction-based quantum devices, such as SQUID and
superconducting qubit.

\end{abstract}
% 74.50.+r  Tunneling phenomena; Josephson effects
% 75.70.-i           Magnetic properties of thin films, surfaces, and interfaces
%74.81.-g            Inhomogeneous superconductors and superconducting systems, including electronic
%                        inhomogeneities

\maketitle

%\narrowtext

\section{Introduction} 
%After nearly a quarter century of the discovery
%of high-temperature cuprate superconductors, there is still no
%consensus on the mechanism for high-$T_c$
%superconductivity~\cite{PWAnderson:97,PALee:06,PWAnderson:07}.
As a remarkable aspect of high-$T_c$ superconductivity, its unique 
characteristics may result from the competition between more than 
one type of order parameter.~\cite{PWAnderson:97,PALee:06,SChakravarty:04} 
Historically, the typical antagonistic relationship between
superconductivity and magnetism has led researchers to avoid using
magnetic elements, such as iron, as potential building blocks of
superconducting (SC) material. 
%The recent discovery of iron-based
%high-$T_c$ superconductor \cite{review10} indicates that our
%fundamental understanding of the origins of superconductivity needs
%significant improvement. 
However, it is now well accepted that the %$d$-wave 
unconventional superconductivity
emerging upon doping is closely related to the antiferromagnetism
(AFM) in parent
compounds~\cite{JZaanen:89,KMachida:89,SAKivelson:03,EDemler:04}.
Recent research theme in high-$T_c$ cuprate community 
centers on how to establish the connection
between antiferromagnetic  (AF) and $d$-wave SC (DSC)
orderings~\cite{orenstein00}, i.e., whether they compete with each
other or coexist microscopically. This issue is a subject of current
discussions. Depending on material details, some compounds show the
coexistence of DSC and AFM~\cite{YSLee:99} while others exhibit
microscopic separation of these two phases~\cite{SHPan:01}.
Engineered heterogeneous systems and multilayered high-$T_c$
cuprates offer a unique setting to study the interplay between DSC
and AFM. In these systems, the disorder effect can be minimized 
significantly with atomically smooth interfaces. 
 Experimentally, no mixing of DSC and AFM was reported in
the heterostructure artificially grown by stacking integer number of
$\mathrm{La_{1.85}Sr_{0.15}CuO_{4}}$ and $\mathrm{La_{2}CuO_{4}}$
layers~\cite{bozovic}. Meanwhile microscopic evidence for the
uniform mixed phase of AFM and DSC in outer $\mathrm{CuO_{2}}$
planes was reported~\cite{mukuda} on a Hg-based five-layered
cuprate. Theoretically, Demler {\em et al.}~\cite{phenomenalogical}
have studied the proximity effect and Josephson coupling in the
SO(5) theory of high-$T_{c}$ superconductors. Depending on the
thickness, the middle antiferromagnetic region could behave like a
superconductor, metal or insulator. On the other hand, it was shown
in the perturbation theory~\cite{perturbative} that the spin
exchange coupling in the insulating AF layer can allow the tunneling
of Cooper pairs.

In this article, we study microscopically a multi-layered system with
two superconductors separated by AF layers. Through self-consistent
mean-field theory and numerical diagonalization, we are able to
solve the Hamiltonian, which allows us to study the interplay
between DSC and AFM in detail. Varying some external parameters,
such as the SC phase difference across the system $\delta\varphi$
and Coulomb interaction $U$ in the AF layers leads to interesting
results about the  interplay.

\begin{figure}[ht]
%\centerline{\psfig{fig1.eps,width=8cm}}
\includegraphics[width=8cm]{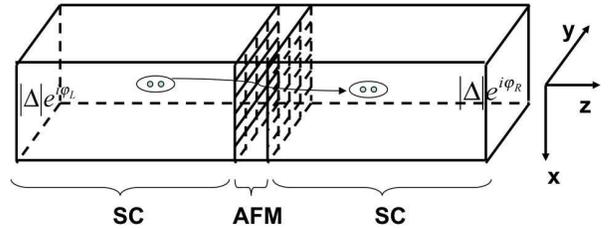}
\caption{Schematic drawing of a multi-layered system. The middle region consists of
AF layers (in X-Y plane) and the two sides are SC layers with
$d$-wave pairing symmetry. The phases of the SC order parameter (or the pairing potential) in the
first (far left) and the last (far right) layer are fixed at
$\varphi_{L}$ and $\varphi_{R}$, respectively.} \label{FIG:Josephson}
\end{figure}

\section{Model and setup} The model system under consideration is
schematically presented in Fig.~\ref{FIG:Josephson}, for which the
Hamiltonian can be written as:
\begin{eqnarray}
H &=& -\sum_{i,j,\sigma}t_{i,j} c_{i,\sigma}^{\dagger} c_{j,\sigma}
+
\sum_{i} U_{i} (n_{i,\uparrow}-\frac{1}{2}) (n_{i,\downarrow}-\frac{1}{2}) \nonumber \\
&& -\frac{1}{2} \sum_{i,j} V_{ij} n_{i}n_{j} -\mu\sum_{i}n_{i}\;, \label{1}
\end{eqnarray}
where $t_{i,j}$ denotes the hopping integral between the nearest
neighbor sites. For in-plane (X-Y plane) hopping, $t_{i,j}=t$, and
for inter-layer (Z direction) hopping, $t_{i,j}=t_{\perp}$.
$c_{i,\sigma}$ ($c_{i,\sigma}^{\dagger}$) is the annihilation
(creation)  operator of electrons on the $i$th lattice site with
spin $\sigma$ ($\sigma=\uparrow, \downarrow$). The quantity
$n_{i,\sigma}=c_{i,\sigma}^{\dagger} c_{i,\sigma}$ is the number
operator on the  $i$th lattice site. Both $U_{i}$ and $V_{ij}$ are
positive. $U_{i}$ indicates the on-site repulsive Coulomb
interaction, which is nonzero \emph{only} in the middle layers, and
$V_{ij}$ describes the in-plane nearest neighbor attractive
interaction, which is nonzero \emph{only} in the SC layers.
%We will not focus on the origin of the nearest neighbor attractive interaction $V$.
We are going to fix the phase of the SC order parameter in the first
(far left) and the last (far right) layer at $\varphi_{L}$ and $
\varphi_{R}$, respectively, which can be introduced by a gauge flux.

When there is no inter-layer coupling, $t_{\perp}=0$, the system is
decomposed into individual two-dimensional (2D) systems. It is known
that in a 2D SC layer, the in-plane nearest neighbor attractive
interaction $V_{ij}$ leads to  a nonzero energy gap $\Delta$
with $d$-wave symmetry, and in a 2D AF layer the on-site repulsive
interaction $U_{i}$ leads to a nonzero $(\pi,\pi)$ spin density wave
(SDW) order. In a genuine cuprate compound, there exists the inter-layer
coupling, which is usually weaker than the in-plane hopping. Because
of the inter-layer coupling, there  arises interesting interplay
between SC layers and AF layers (see Fig.~\ref{FIG:Josephson}).

The strategy of solving the Hamiltonian is briefly summarized as
follows: Due to the translational symmetry in the 2D X-Y plane, the
3D problem can be decomposed into a 2D (X-Y) plus 1D (Z) problem. 
In the X-Y plane, $\vec{k}= (k_{x}, k_{y})$ is a good quantum number, where
$k_{x}=\frac{1}{2}(\frac{n_{x}}{N_{x}}-\frac{n_{y}}{N_{y}})\pi$,
$k_{y}=\frac{1}{2}(\frac{n_{x}}{N_{x}}+\frac{n_{y}}{N_{y}})\pi$, and 
$n_{x}=-\frac{1}{2} N_{x}$, $-\frac{1}{2} N_{x}+1$, $\cdots$, $\frac{1}{2} N_{x}-1$,
$n_{y}=-\frac{1}{2} N_{y}$, $-\frac{1}{2} N_{y}+1$, $\cdots$, $\frac{1}{2} N_{y}-1$.
In order to simplify the Hamiltonian, we adopt the mean-field
approximation, which leads to the following Bogoliubov-de Gennes
equation~\cite{PGdeGennes:1965}:
\begin{widetext}
\begin{equation}
\sum_{m} \left[
\begin{array}{cccc}
\xi_{\vec{k}n} \delta_{nm}+t_{nm} & -\frac{U_n}{2}M_{n}\delta_{nm} &\Delta_{\vec{k},n}\delta_{nm}&0 \\
 -\frac{U_n}{2}M_{n}\delta_{nm} & \xi_{\vec{k}+\vec{Q},n} \delta_{nm}+t_{nm}  &0 & \Delta_{\vec{k}+\vec{Q},n} \delta_{nm} \\
\Delta_{\vec{k},n}^{\ast}\delta_{nm} & 0&-\xi_{-\vec{k}n} \delta_{nm}-t_{nm}^{\ast}  &-\frac{U_n}{2}M_{n}\delta_{nm}  \\
 0& \Delta_{\vec{k}+\vec{Q},n}^{\ast}\delta_{nm} &-\frac{U_n}{2}M_{n}\delta_{nm}  &-\xi_{-(\vec{k}+\vec{Q})n} \delta_{nm}-t_{nm}^{\ast}  \\
\end{array}
\right]\left[
\begin{array}{c}
u_{\vec{k},m,\uparrow}^{\alpha}  \\
u_{\vec{k}+\vec{Q},m,\uparrow}^{\alpha}\\
v_{\vec{k},m,\downarrow}^{\alpha}  \\
v_{\vec{k}+\vec{Q},m,\downarrow}^{\alpha}
\end{array}
\right]
= E_{\alpha}
\left[
\begin{array}{c}
u_{\vec{k},n,\uparrow}^{\alpha}  \\
u_{\vec{k}+\vec{Q},n,\uparrow}^{\alpha}\\
v_{\vec{k},n,\downarrow}^{\alpha}  \\
v_{\vec{k}+\vec{Q},n,\downarrow}^{\alpha}
\end{array}
\right]\;, %\nonumber
\end{equation}
\end{widetext}
where $m$ and $n$ label the layer number; $\vec{k}$ and
$\vec{k}+\vec{Q}$ with $\vec{Q}=(\pi,\pi)$ denote the wave vectors
in the first Brillouin zone of the X-Y plane,
$t_{mn}=t_{\perp}(\delta_{n,m+1}+\delta_{n,m-1})$ describes the
nearest neighbor inter-plane hopping, $\xi_{\vec{k}n}=-2 t (\cos
k_{x}+ \cos k_{y})-\mu$ is the normal metal dispersion relation,
$\mu$ is the chemical potential, the variables $U_n$ is equal to $U$
in the AF layers and zero otherwise while $V_n$ is equal to $V$ in
the SC layers and zero otherwise. The superconducting pairing potential 
 $\Delta_{\vec{k},n}=V_{n}\Psi_{n}(\cos k_{x}-\cos k_{y})/2$, where $\Psi_{n}$ is the superconducting 
 pairing wavefunction. Obviously, the  superconducting pairing potential, which is the SC order parameter
has the same phase as that of the superconducting 
 pairing wavefunction. $M_{n}=\left\langle
n_{n,\uparrow}\right\rangle - \left\langle
n_{n,\downarrow}\right\rangle$ describes the SDW in the $n$-th
layer. The average electron number $\left\langle
n_{n,\uparrow}\right\rangle$, $\left\langle
n_{n,\downarrow}\right\rangle$ and the SC pairing wavefunction
$\Psi_{n}$ can be determined self-consistently through iteration
over the following relation:
\begin{subequations}
\begin{eqnarray}
\Psi_{n}= \frac{2}{N} \sum_{\vec{k},\alpha} \frac{(\cos k_{x}-\cos k_{y})}{2} g(u,v) \tanh{\frac{\beta E_{\alpha}}{2}}\;,\\
\left\langle n_{n,\uparrow}\right\rangle= \frac{1}{N}
\sum_{\vec{k},\alpha} \left| u_{\vec{k},n,\uparrow}^{\alpha} +
u_{\vec{k}+\vec{Q},n,\uparrow}^{\alpha} \right |^{2} f(E_{\alpha})\;,\\
\left\langle n_{n,\downarrow}\right\rangle= \frac{1}{N}
\sum_{\vec{k},\alpha} \left| v_{\vec{k},n,\downarrow}^{\alpha} +
v_{\vec{k}+\vec{Q},n,\downarrow}^{\alpha} \right |^{2}
f(-E_{\alpha})\;,
\end{eqnarray}
\end{subequations}
where $N=N_{x}\times N_{y}$ is the sites number in a 2D plane.
$\vec{k}$ samples half of the first Brillouin zone.
$g(u,v)=u_{\vec{k},n,\uparrow}^{\alpha}
v_{\vec{k},n,\downarrow}^{\alpha, \ast} -
u_{\vec{k}+\vec{Q},n,\uparrow}^{\alpha}
v_{\vec{k}+\vec{Q},n,\downarrow}^{\alpha, \ast}$, and
$f(E_{\alpha})=1/(1+e^{\beta E_{\alpha}})$ is the Fermi
distribution. Through this simplification, the multi-layered system
is then solved numerically. We are especially interested in the
interplay between the DSC and the AFM, which is characterized by the
SDW $M_{n}$ and the SC pairing wavefunction $\Psi_{n}$. We will focus
on zero temperature $T=0$. The size of X-Y plane is
$N_{x}=N_{y}=40$. The total layer number is $N_{z}=22$. The two
middle layers are AF layers ($U_{n}=U$, $V_{n}=0$, for $n=11$, $12$), and the
rest are SC layers ($U_{n}=0$, $V_{n}=V$, for $n=1$, $2$, $\cdots$, $10$, and $n=13$, $\cdots$, $22$). For simplicity, we
choose $t=1$, $t_{\perp}=0.1$, $V=1.5$, and  $\mu=0$. We will vary
the control  parameters, such as  $U$ and  $\delta\varphi=
\varphi_{L}-\varphi_{R}$, to study the competition between DSC and
AFM.  When the phase difference $\delta\varphi$ is introduced, there
is a current flowing across the system and its expression is given
by:
\begin{eqnarray}
I &= & 2\frac{t_{\perp}e}{N}  \sum_{\vec{k},\alpha} \mathrm{Im} \biggl{\{} \left [ u_{\vec{k},n,\uparrow}^{\alpha, \ast} u_{\vec{k},n+1,\uparrow}^{\alpha} +  \left (\vec{k} \leftrightarrow  \vec{k}+\vec{Q}\right) \right] f_{+} \nonumber \\
&&   + \left[ v_{\vec{k},n,\downarrow}^{\alpha} v_{\vec{k},n+1,\downarrow}^{\alpha, \ast} +  v_{\vec{k}+\vec{Q},n,\downarrow}^{\alpha} v_{\vec{k}+\vec{Q},n+1,\downarrow}^{\alpha, \ast}\right] f_{-}  \biggr{\}}\;,
\end{eqnarray}
where $f_{\pm} = f(\pm E_{\alpha})$.
%Having introduced the model and
%the method for solving it, in the following, we will do the
%simulation and study the interplay through the numerical result.

\section{Influence of antiferromagnetism on superconductivity} In the
model introduced above, the change of the Coulomb interaction $U$ in
the middle layers can drive a metal-Mott insulator phase transition
at a certain value $U_{c}$. Correspondingly, the multi-layered system changes from a superconductor/normal
metal/superconductor (SNS) weak link \cite {bardeen72} to a
superconductor/insulator/superconductor (SIS) Josephson junction
\cite{josephson}. The studies of SNS and SIS junctions with static
potential barrier have been well
documented~\cite{bardeen72,josephson,MTinkham:75}. Here we start
from a microscopic model and drive the middle layers to change from
one electronic state into another by tuning the on-site Coulomb
interaction $U$. We are interested in how the SC pairing wavefunction and
the current across the junction change with the Coulomb interaction
$U$. We plot the current as a function of the phase difference
$\delta\varphi$ in Fig. 2. It can be seen that below a threshold
value of $U_{c} \approx 0.8$, the current is a piecewise periodic
function. In each periodic region, it varies linearly with
$\delta\varphi$. This agrees with the result obtained in
Ref.~\onlinecite{bardeen72} for a junction with a normal metal. We also find that
%when we vary $U$, as long as it is below the critical value $U_{c}$, the current is not influenced by the change of $U$. This means that the superconductor is not influenced by the AFM. However,
when the Coulomb interaction is larger than $U_{c}$, the current as
a function of $\delta \varphi$ changes gradually with the increase of
$U$. When $U$ is in the range $0.8<U<1.5$, the current shows a shape intermediate
between a piecewise linear function and a sinusoidal
function. Similar results have been observed in Refs.~\cite{bardeen72, phenomenalogical,EDemler:04}
when one varies the temperature or the thickness of the middle layers 
instead of the the Coulomb interaction $U$. When $U\approx 2.0$, the current is well approximated by a
sinusoidal function of $\delta\varphi$ (see Fig. 2), which is a
typical feature of the dc Josephson junction (JJ)
current~\cite{josephson}. The magnitude of current decreases rapidly
with the increase of $U$. 

\begin{figure}[t]
%\centerline{\psfig{fig1.eps.eps,width=8cm}}
\includegraphics[width=7cm, angle=0]{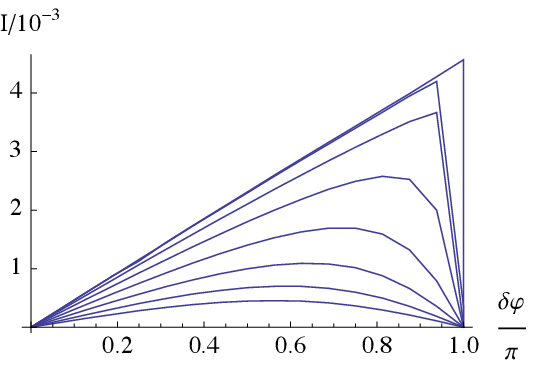}
\includegraphics[width=7cm, angle=0]{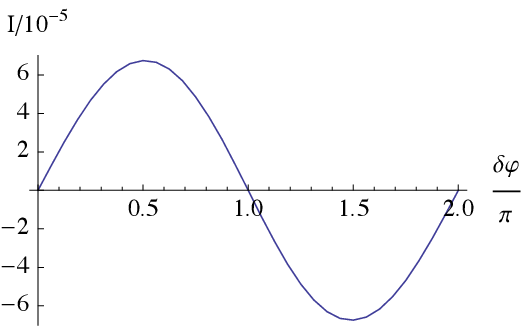}
\caption{Current across the junction as a function
of the phase difference $\delta \varphi$ of two superconductors for
different $U$. Here $V=1.5$, $t_{\perp}=0.1$, $N_{x}=N_{y}=40$,
$N_{z}=22$. Up panel: $U=0$, $0.1$,  $0.2$, $\cdots$, $1.5$, and
$\delta \varphi=0 \sim \pi$. Down panel: $U=2.0$ and $\delta
\varphi=0 \sim 2\pi$.} \label{FIG:current}
\end{figure}
\begin{figure}[t]
%\centerline{\psfig{fig1.eps.eps,width=8cm}}
\includegraphics[width=6cm, angle=-90]{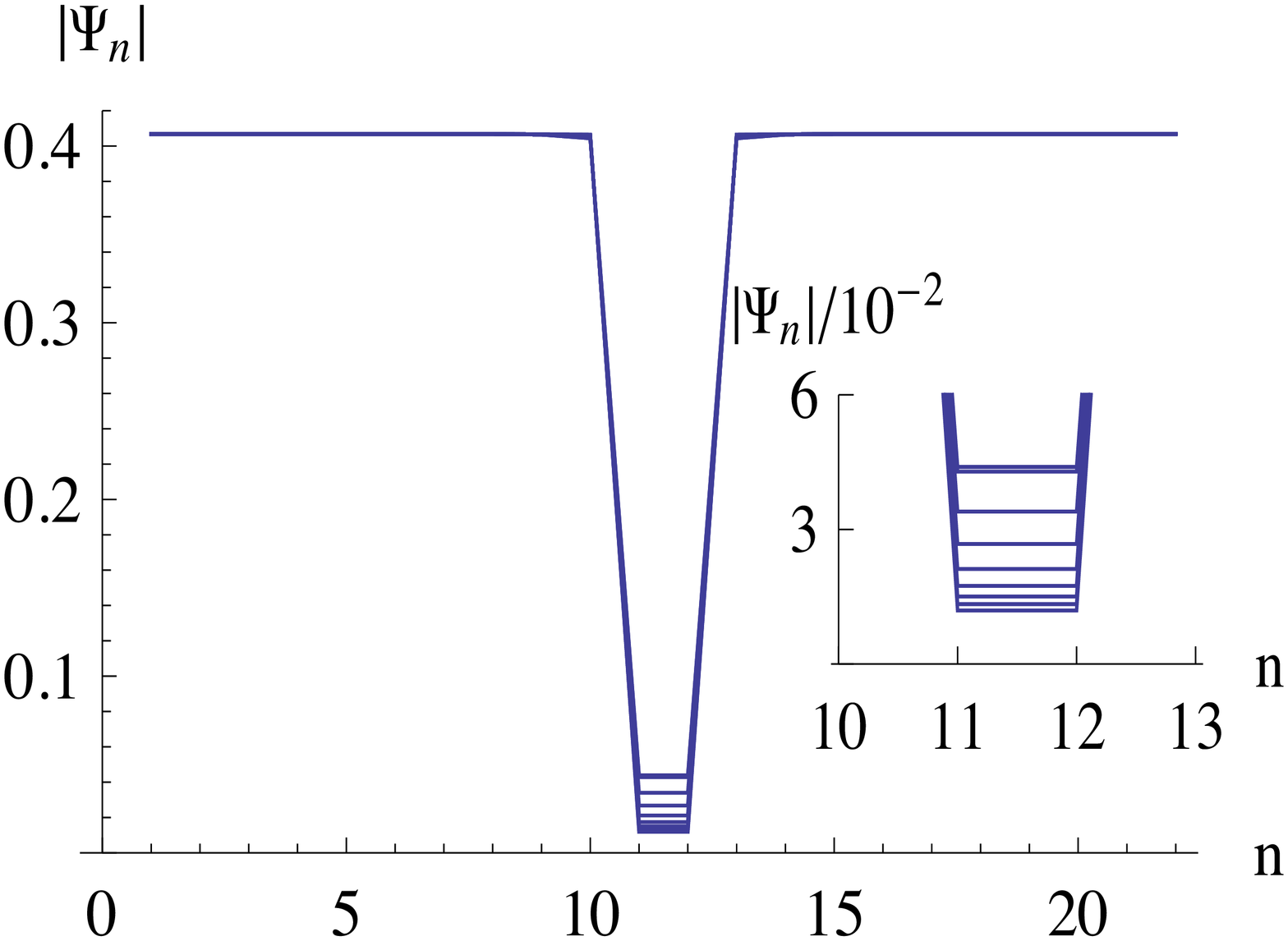}
\includegraphics[width=6cm, angle=-90]{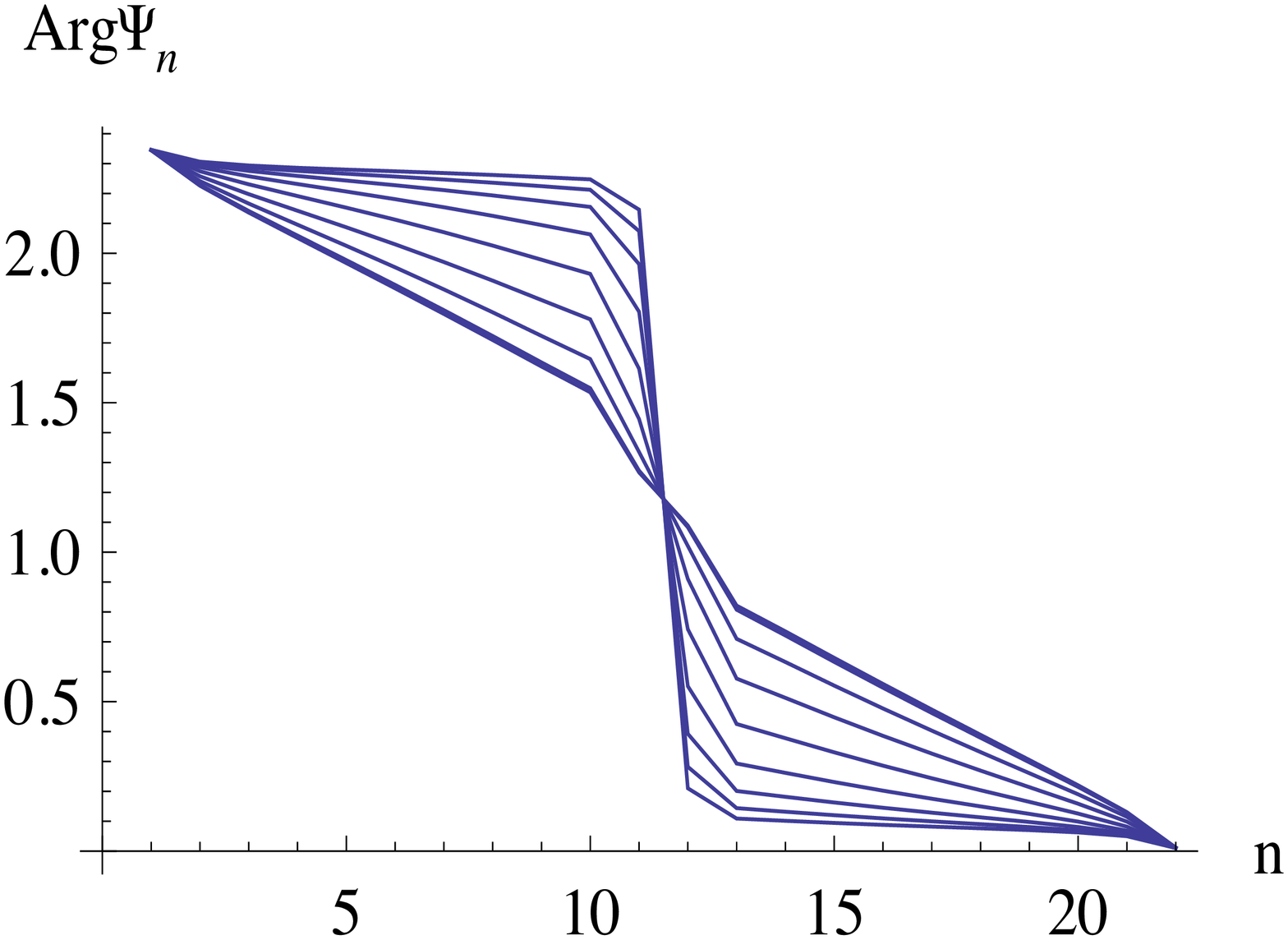}
\caption{Absolute value (Up panel) and argument
(Down panel) of superconducting pairing wavefunction $\Psi_{n}$ as a function of
layer number. The phase of SC pairing wavefunction is the
same as that of the pairing wavefunction. Here $V=1.5$, $t_{\perp}=0.1$, $N_{x}=N_{y}=40$,
$N_{z}=22$, $\delta \varphi=3\pi/4$. $U$ is fixed at different
values: $U=0$, $0.1$,  $0.2$, $\cdots$, $1.5$. }
\label{FIG:orderparameter}
\end{figure}

In order to have a better understanding of the influence of
AFM on DSC, we plot the absolute value and the phase of the SC 
 pairing wavefunction in Fig.~\ref{FIG:orderparameter}. It can be seen that when
$U$ is small, there is a nonzero paring potential in the middle
layers due to the proximity effect, and the two superconductors are
weakly linked by the normal metallic layers. In this weak link
regime, the phase of the SC pairing wavefunction is almost linear in
layer number $n$. This means that when the middle layers are in the
normal metallic state, it has little influence on the SC layers on
two sides. However, when $U$ is larger than $U_{c} \approx 0.8$, SDW
develops, and the middle layers become Mott insulator. Quantum
mechanically, Cooper pairs can still tunnel through the SDW region
when $U$ is not very large. We call this intermediate  $U$ regime as
the tunneling regime. From the weak link regime to tunneling regime,
%a notable change in the phase distribution of the SC order parameter
%occurs.
the phase of the SC paring wavefunction changes gradually from a
(nearly) linear function to a step function of the layer number (see
Fig.~\ref{FIG:orderparameter}). This corresponds to  the observation
that the current shape changes from being piecewise linear to
sinusoidal. Therefore, when $U$ is very large, the current-phase
relation becomes identical to the famous Josephson
relation~\cite{josephson,feynman}. Here for the first time, we
establish the connection between the profile of the SC phase
distribution (Fig.~\ref{FIG:orderparameter}) and the current-phase dependence (Fig.~\ref{FIG:current}), which we believe is
rather intriguing.

When one continues to increase $U$, the middle layers becomes more
opaque and  the superconducting pairing wavefunction is completely suppressed in
this region. As a result, the tunneling of Cooer pairs will be
completely quenched, and the two comprising superconductors become
decoupled.

We also would like to mention that when the middle layers are in the
normal metallic state, the current across the junction is a
mesoscopic effect~\cite{JXZhu:10}. The current is proportional to
the phase gradient of the SC pairing wavefunction. That is, when one
increases the  number of SC layers, the current will decrease and
finally vanish. However, when the middle layer is in the AF state,
the whole system becomes a JJ. The current of a JJ is no longer a
mesoscopic effect. It is completely determined by the phase
difference of the SC pairing wavefunction on both ends of the system.
Therefore,  it will not decrease with the increase of the  number of
the SC layers. In this sense we say that AFM insulator layers enhance supercurrent.

\section{Influence of superconductivity on antiferromagnetism} 
\begin{figure}[t]
%\centerline{\psfig{fig1.eps.eps,width=8cm}}
\includegraphics[width=7cm, angle=0]{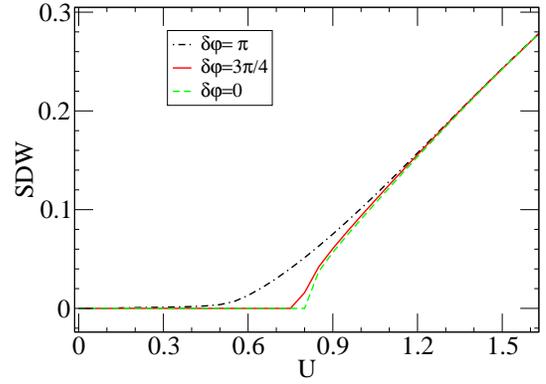}
\caption{(Color online)  SDW in the middle layer ($n=11$) as a
function of the Coulomb interaction for different fixed $\delta
\varphi$. Here $U=0 \sim 1.5$, $V=1.5$, $t_{\perp}=0.1$,
$N_{x}=N_{y}=40$, $N_{z}=22$, $\delta \varphi=0$ (solid), $3\pi/4$
(dashed), and $\pi$ (dot dashed).} \label{FIG:SDW}
\end{figure}

In the
above discussion, we find the profound influence of the Coulomb
interaction $U$ (and hence AFM) of the middle layers on the DSC. We
now study the back-action of DSC on AFM. We will fix the phase
difference $\delta\varphi$ at different values, and see if the SDW
in the middle layers changes with $\delta\varphi$. We plot in
Fig.~\ref{FIG:SDW}  the SDW as a function of $U$ for various values
of  $\delta \varphi$. Clearly, the SDW in the middle
layers is influenced by $\delta \varphi$ in certain range of $U$.
When $\delta \varphi =\pi$, the critical value $U_{c}$, after which
the SDW develops, is much smaller than in the case of $\delta
\varphi=0$, though the current is zero in both cases. This is very
similar to the result in Ref.~\onlinecite{phenomenalogical} that the
middle AF layers will be influenced by the current fed into the
junction. When the Coulomb interaction is very weak, the SDW does
not develop for $\delta \varphi \neq \pi$, and the middle layers are
in the normal metallic state. In contrast, when the Coulomb interaction
is very strong, e.g. $U>1.2$,  the AFM is robust against the phase
difference across the junction and strongly suppresses the
pairing wavefunction in the middle region (see
Fig.~\ref{FIG:orderparameter}). Only when $U$ is small, is the
emergence of the SDW sensitive to the phase difference.

%Let us recall that in a 2D Hubbard model ($t_{\perp}=0$), there is no influence of DSC on AFM. 
When switching off the interlayer coupling ($t_{\perp}=0$), the system is decoupled
into 2D systems. The DSC on two sides will not influence the AFM in the middle layer.  
The middle layer is described by a 2D Hubbard model, in which the SDW will develop for an
infinitesimal $U$. In the present model ($t_{\perp}\neq0$), when $\delta \varphi=0$,
the SDW in the middle layers is suppressed due to the inter-layer
coupling, which weakens the perfect nesting at $\vec{Q}=(\pi,\pi)$.
The SDW will not develop until $U$ is larger than a threshold value
$U_{c}$ (see Fig.~\ref{FIG:SDW}). Hence both $t_{\perp}$ and
$\delta\varphi$ will influence the AFM. The results in
Fig.~\ref{FIG:SDW} indicate that the AFM shows a crossover behavior
for $\delta \varphi =\pi$ rather than a criticality as obtained for
$\delta \varphi=0$.  For $\delta \varphi \in (0, \pi)$, the SDW
critical point $U_{c}$ is less than that for $\delta \varphi=0$.
 This result is in agreement with that in Ref.~\onlinecite{phenomenalogical}.

\begin{figure}[t]
%\centerline{\psfig{fig1.eps.eps,width=8cm}}
\includegraphics[width=7cm, angle=0]{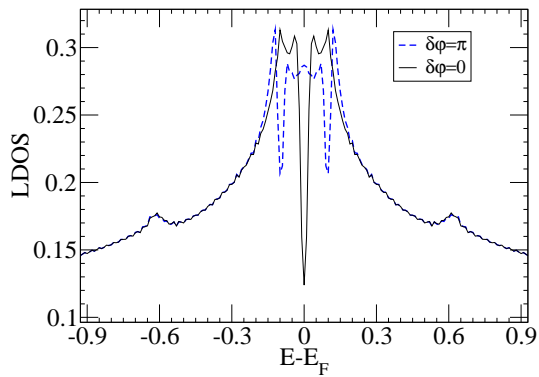}
\caption{(Color online)  LDOS of the middle layer ($n=11$)
$\rho_{11,\uparrow}(E)$. Here $U=0$, $V=1.5$, $t_{\perp}=0.1$,
$N_{x}=N_{y}=40$, $N_{z}=22$, $\delta \varphi=0$, and $\pi$.}
\label{FIG:LDOS}
\end{figure}

In order to have a better understanding of the influence of the
DSC on the AFM in the middle layers, we plot in Fig.~\ref{FIG:LDOS}
the local density of state (LDOS), as given by
\begin{subequations}
\begin{eqnarray}
\rho_{n,\uparrow}(E)&=&\frac{1}{N}  \sum_{\vec{k},\alpha} \left| u_{\vec{k},n,\uparrow}^{\alpha}  +u_{\vec{k}+\vec{Q},n,\uparrow}^{\alpha} \right|^{2} \left[ -\frac{\partial f(\omega)}{\partial \omega}\right],\\
\rho_{n,\downarrow}(E)&=&\frac{1}{N}  \sum_{\vec{k},\alpha} \left| v_{\vec{k},n,\downarrow}^{\alpha}  +v_{\vec{k}+\vec{Q},n,\downarrow}^{\alpha} \right|^{2} \left[ -\frac{\partial f(\omega')}{\partial \omega'}\right],
\label{5}
\end{eqnarray}
\end{subequations}
where $\omega=E-E_{\alpha}$ and $\omega'=E+E_{\alpha}$. From
Fig.~\ref{FIG:LDOS}, we can see a sharp intensity decay of LDOS at
$E_{F}$ when $\delta \varphi =0$. When $\delta \varphi =\pi$, the
LDOS exhibits a peak structure at $E_{F}$. Since the development of
the SDW is also determined by the normal-state DOS at low energies,
it explains the different SDW critical behavior between the cases
with $\delta \varphi =\pi$ and $\delta \varphi = 0$.

\section{Conclusion and discussion} We have studied the interplay
between the DSC and AFM in a multi-layered system. It is revealed
that the AF layers have profound influence on the DSC. A metal-Mott
insulator phase transition makes the multi-layer system change from
a SNS to a SIS junction. We have also established the connection
between the profile of the SC phase distribution and the
current-phase dependence. We find that when the barrier of the middle layers is
not high enough (metal or semiconductor regime), the middle layers cannot maintain a step-function-like 
phase of two superconductors on two sides. Accordingly, the current
across the junction will deviate from a sinusoidal function of the phase difference.
When the barrier of the middle layer is high (deep insulator regime), as predicted by
Josephson \cite{josephson} the phase across the junction is a step function, and the current
is a sinusoidal function of the phase difference.
Meanwhile, the phase difference across the
system will dramatically influence the development of SDW in the
middle layers. %When the phase difference $\delta \varphi$ is equal
%to $\pi$, the SDW in the middle layer develops earlier than any
%other $\delta \varphi$ \cite{phenomenalogical}.
The results from our
simulations may shed new light on material engineering. %the mechanism of high-$T_{c}$ superconductivity. 
The DSC and AFM can coexist in a single layer,
but they also compete with each other.  Another important insight
from the present study is the microscopic model of JJ. JJ has been
widely used in a lot of quantum devices, such as in SQUID for
ultra-sensitive magnetic field measurement and superconducting
qubit~\cite{qubit} for quantum computing. Our study provides a
microscopic theory for JJ, hence has possible applications in
building adaptive JJ based devices by engineering the middle layers
of the JJ. Finally, with the development of experimental techniques,
it is now possible to synthesize heterogeneous systems with layer by
layer at atomic scale~\cite{threelayer}. %This kind of material
%engineering provides a powerful method for understanding the
%fundamental physics of high-$T_{c}$ superconductivity.
We expect that our results can be experimentally verified, as some
experiments on engineered  material  have been
reported~\cite{threelayer,bozovic,mukuda}.

{\bf Acknowledgments:} We thank A. V. Balatsky, Quanxi Jia, A. J. Taylor, and
S. Trugman for useful discussions. This work was carried out under the auspices
of the National Nuclear Security Administration of the U.S. DOE at
LANL under Contract No. DE-AC52-06NA25396, the LANL LDRD-DR Project X96Y,
and the U.S. DOE Office of Science.

\end{document}